\documentclass[reprint,amsmath,amssymb,aps,prl,twocolumn]{revtex4-2}

\usepackage{graphicx}
\usepackage{dcolumn}
\usepackage{bm}
\usepackage{hyperref}
\usepackage[mathlines]{lineno}
\usepackage{xcolor}
\newcommand{\overbar}[1]{\mkern 1.5mu\overline{\mkern-1.5mu#1\mkern-1.5mu}
\mkern 1.5mu}
\newcommand{\rme}{\mathrm{e}}
\newcommand{\tr}{\mathrm{tr}\,}

\begin{document}
\title{Structure constants in \texorpdfstring{$\mathcal{N}=2$}{}
superconformal quiver theories at strong coupling
\texorpdfstring{\\}{} and holography}
\thanks{This work is partially supported by the MIUR PRIN Grant 2020KR4KN2 ``String Theory as a bridge between Gauge Theories and Quantum Gravity".}

\author{Marco Bill\`o$\,^{a,b}$}
\author{Marialuisa Frau$\,^{a,b}$}
\author{Alberto Lerda$\,^{c,b}$}
\author{Alessandro Pini$\,^{b}$}
\author{Paolo Vallarino$\,^{a,b}$}
\email{billo,frau,lerda,apini,vallarin@to.infn.it}

\affiliation{$^{a}$
	Universit\`a di Torino, Dipartimento di Fisica,
	Via P. Giuria 1, I-10125 Torino, Italy}

\affiliation{$^{b}$
	Istituto Nazionale di Fisica Nucleare - sezione di Torino,
	Via P. Giuria 1, I-10125 Torino, Italy}

\affiliation{$^{c}$
	Universit\`a del Piemonte Orientale,
	Dipartimento di Scienze e Innovazione Tecnologica,
	Viale T. Michel 11, I-15121 Alessandria, Italy }

\begin{abstract}
In a four-dimensional $\mathcal{N}=2$ superconformal quiver theory with gauge group $\mathrm{SU}(N)\times\mathrm{SU}(N)$ and bi-fundamental matter, we analytically obtain the exact strong-coupling behavior of the normalized 3-point correlators of single-trace scalar operators in the large-$N$ limit using localization techniques.
We then obtain the same strong-coupling behavior from the holographic dual using the AdS/CFT correspondence at the supergravity level. This agreement confirms the validity of the analytic 
strong-coupling results and of the holographic correspondence in a non-maximally supersymmetric set-up in four dimensions.
\end{abstract}


\maketitle

\section{Introduction}
\label{sec:intro}
\begin{figure}[h]
	\includegraphics[width=0.27\columnwidth]{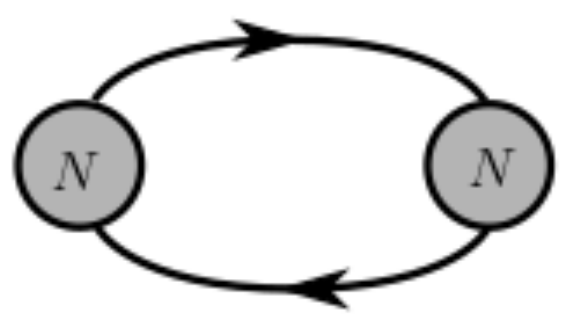}
	\caption{\label{fig:model}The quiver diagram of the $\mathrm{SU}(N)\times\mathrm{SU}(N)$ superconformal theory. Each circle denotes a SU($N$) factor and the oriented lines represent the bi-fundamental hypermultiplets.}
\end{figure}

We consider a supersymmetric quiver theory in four dimensions with gauge group 
$\mathrm{SU}(N)\times\mathrm{SU}(N)$ and bi-fundamental matter, schematically represented in
Fig.\,\ref{fig:model}. 
This model, which arises as a $\mathbb{Z}_2$ orbifold of $\mathcal{N}=4$ Super Yang-Mills (SYM) theory, has $\mathcal{N}=2$ supersymmetry, is conformally invariant at the
quantum level and admits a holographic dual description in terms of Type II B string theory on 
AdS$_5\times S^5/\mathbb{Z}_2$ \cite{Kachru:1998ys,Gukov:1998kk}. As such it is
one of the simplest four-dimensional $\mathcal{N}=2$ theories to explore the strong-coupling regime and probe the holographic correspondence in a non-maximally supersymmetric set-up
(see also \cite{Zarembo:2020tpf}).

With this aim, we study 2- and 3-point functions
of protected gauge-invariant operators defined as
\begin{eqnarray}
U_k(x)&=\frac{1}{\sqrt{2}}\big[\mathrm{tr}\,\phi_0(x)^k+\mathrm{tr}\,\phi_1(x)^k
\big]~,\label{U}\\
T_k(x)&=\frac{1}{\sqrt{2}}\big[\mathrm{tr}\,\phi_0(x)^k-\mathrm{tr}\,\phi_1(x)^k
\Big]~.\label{T}
\end{eqnarray}
Here $k$ is an integer $\geq 2$ and $\phi_{0,1}$ are the chiral scalar fields of the
adjoint vector multiplets in the two nodes of the quiver. Replacing $\phi_{0,1}$ with their complex
conjugates $\overbar{\phi}_{0,1}$ yields the anti-chiral operators $\overbar{U}_k$ and $\overbar{T}_k$. All are primary operators with conformal dimension $\Delta=k$ and charge
$Q=\pm k$ in the chiral or anti-chiral case. We call the operators (\ref{U}) and (\ref{T}) untwisted 
($U$) and twisted ($T$), since they are, respectively, even and odd under the $\mathbb{Z}_2$
symmetry exchanging the nodes of the quiver.

Conformal invariance, charge conservation and $\mathbb{Z}_2$ symmetry fix the form of the 2-point functions to be
\begin{eqnarray}
\big\langle U_k(x)\,\overbar{U}_k(y)\big\rangle
&=\displaystyle{\frac{G_{U_k}}{|x-y|^{2k}}}
~,\label{UU}\\
\big\langle T_k(x)\,\overbar{T}_k(y)\big\rangle
&=\displaystyle{\frac{G_{T_k}}{|x-y|^{2k}}}~,
\label{TT}
\end{eqnarray}
where the coefficients $G_{U_k}$ and $G_{T_k}$ depend on $k$, $N$ and the 't Hooft coupling $\lambda$.
Also the 3-point functions are constrained by the symmetries of the theory. Here we will consider the following correlators:
\begin{eqnarray}
\big\langle U_k(x)\,U_\ell(y)\,\overbar{U}_p(z)\big\rangle
&=\displaystyle{\frac{G_{U_kU_\ell\overbar{U}_p}}{|x-z|^{2k}\,|y-z|^{2\ell}}}~,
\label{UUU}\\
\big\langle U_k(x)\,T_\ell(y)\,\overbar{T}_p(z)\big\rangle
&=\displaystyle{\frac{G_{U_kT_\ell\overbar{T}_p}}{|x-z|^{2k}\,|y-z|^{2\ell}}}~,
\label{UTT}
\end{eqnarray}
with the understanding that $p=k+\ell$ for charge conservation. 
Again the coefficients in the numerators are functions of $N$, $\lambda$ and of the conformal dimensions. One could consider also the conjugate 3-point functions where chiral and anti-chiral operators are exchanged, as well as $\big\langle T_k(x)\,T_\ell(y)\,\overbar{U}_p(z)\big\rangle$ and their conjugates. For simplicity in this Letter we focus on the above cases.

The coefficients $G$ in (\ref{UU})--(\ref{UTT}) are sensitive to the 
normalization of the operators. To remove such dependence we define the structure constants
\begin{eqnarray}
C_{U_kU_\ell\overbar{U}_p}&=\displaystyle{\frac{G_{U_k U_\ell \overbar{U}_p}}{\sqrt{G_{U_k}\,
G_{U_\ell} G_{U_p}}}}~,\label{CUUUdef}\\
C_{U_kT_\ell\overbar{T}_p}&=\displaystyle{\frac{G_{U_k T_\ell \overbar{T}_p}}{\sqrt{G_{U_k}\,
G_{T_\ell} G_{T_p}}}}~,\label{CUTTdef}
\end{eqnarray}
which, together with the spectrum of conformal operators, are part of the conformal field theory
data.
We study these structure constants in the large-$N$ 't Hooft limit. Using supersymmetric
localization, we analytically obtain the exact
dependence on $\lambda$ and predict the 
strong-coupling behavior, which is then obtained also with a holographic calculation using the Anti de Sitter/Conformal Field Theory (AdS/CFT) correspondence 
\cite{Maldacena:1997re,Gubser:1998bc,Witten:1998qj} \footnote{This Letter contains the main ideas and results, while the derivation and the extension to quiver theories with an arbitrary number of nodes are contained in \cite{Billo:2022fnb}.}.

\section{Localization results}
\label{sec:local}
The flat-space correlators discussed above can be conformally mapped to correlators defined on a 4-sphere. 
These, in turn, can be evaluated \cite{Baggio:2014sna,Gerchkovitz:2016gxx,Billo:2017glv,Beccaria:2020hgy,Galvagno:2020cgq,Beccaria:2021hvt,Billo:2021rdb,Billo:2022xas} in terms of a matrix model determined by localization techniques \cite{Pestun:2007rz}. 

The matrix model has the quiver structure of Fig.\,\ref{fig:model}, with two Hermitean $N\times N$ matrices $a_0$ and $a_1$ defined in the two nodes. Neglecting instanton contributions which are exponentially suppressed in the planar limit, its partition function is
\begin{equation}
	\label{Zmat}
		\mathcal{Z} = \int da_0\, da_1\, \rme^{-\tr a_0^2 - \tr a_1^2 - S_{\mathrm{int}}}
		\equiv \big\langle \rme^{- S_{\mathrm{int}}}\big\rangle_0~,
\end{equation}	 
where $\langle ~ \rangle_0$ indicates the vacuum expectation value with respect to the Gaussian measure, and $S_{\mathrm{int}}$ is a perturbative series in $\lambda$ given 
in Eq.\,(C.5) of \cite{Billo:2021rdb}.
The gauge theory operators $U_k$ and $T_k$ in (\ref{U}) and (\ref{T}) are represented in the matrix model, respectively, by $\mathcal{O}^+_{k}$ and $\mathcal{O}^-_{k}$ with 
\begin{equation}
	\mathcal{O}^\pm_k 
	={\textstyle{\frac{1}{\sqrt{2}}}}\!:\!\big(\mathrm{tr}\,a_0^k \pm \mathrm{tr}\,a_1^k
	\big)\!:~,
	\label{UTmat}
\end{equation}
where the normal-ordering $:\,:$ means that one has to subtract the contractions 
with all operators of lower conformal dimension. As proven in \cite{Baggio:2014sna,Billo:2022xas}, in the large-$N$ limit this amounts to perform a Gram-Schmidt diagonalization within the set of single-trace operators only. Also the anti-chiral operators $\overbar{U}_k$ and $\overbar{T}_k$ are
represented by $\mathcal{O}^{\pm}_k$, and thus the coefficients $G$ in the 2- and 3-point functions can be computed by evaluating the vacuum expectation value of products of such operators. For example,
\begin{equation}
	G_{U_kT_\ell\overbar{T}_p}=
	\big\langle \mathcal{O}^+_k\,\mathcal{O}^-_\ell\,\mathcal{O}^-_p\big\rangle
	\,\equiv\,\frac{1}{\mathcal{Z}} \big\langle \mathcal{O}^+_k\,\mathcal{O}^-_\ell\,\mathcal{O}^-_p
	\, \rme^{-S_{\mathrm{int}}}\big\rangle_0~.
	\label{type11}
\end{equation}
In \cite{Billo:2021rdb} it has been shown that the interaction action $S_{\mathrm{int}}$ only contains the twisted operators $\mathcal{O}^-_k$. This allowed us to evaluate the partition function as
\begin{equation}
	\mathcal{Z}=\det\big(1-\mathsf{X}\big)^{-\frac{1}{2}}~,
\end{equation}
where $\mathsf{X}$ is an infinite matrix whose elements are
\begin{equation}
	\mathsf{X}_{i,j}=-8(-1)^{\frac{i+j+2ij}{2}}\!\sqrt{i j}\!\!\int_0^\infty\!\!
	\!\!\!\frac{dt\,\,\mathrm{e}^t}{t\,(\mathrm{e}^t-1)^2}
	J_{i}\Big(\frac{t\sqrt{\lambda}}{2\pi}\Big) 
	J_{j}\Big(\frac{t\sqrt{\lambda}}{2\pi}\Big).
	\label{X}
\end{equation}
Here $J_i$ is the Bessel function of the first kind and the indices $i$ and $j$ are both even or both odd.
Remarkably, using the full Lie Algebra approach developed in \cite{Billo:2017glv,Beccaria:2020hgy}, one can show that at large $N$ also the expectation values as the one in (\ref{type11}) can be written in closed form in terms of the matrix $\mathsf{X}$. This means that their full dependence on 
$\lambda$ is known through integrals of Bessel functions.

The correlators involving only untwisted operators, which do not appear in $S_{\mathrm{int}}$, are actually $\lambda$-independent. Indeed, at large $N$ we
find
\begin{eqnarray}
	&\hspace{-0.8cm} G_{U_k} =k\,\big(\frac{N}{2}\big)^k\,\equiv\,\mathcal{G}_k~,
	\label{GUloc}\\[1mm]
	&G_{U_k,U_\ell,\overbar{U}_{p}}  =\frac{k\,\ell\,p}{2\sqrt{2}}\,\big(\frac{N}{2}\big)^{\frac{k+\ell+p}{2}-1}\,\equiv\, 
	\mathcal{G}_{k,\ell,p}
	\label{GUUUloc}
\end{eqnarray}
for all values of $\lambda$. From this it easily follows that
\begin{equation}
C_{U_k U_\ell \overbar{U}_{p}} 
=\frac{\sqrt{k\,\ell\,p}\phantom{\big|}}{\sqrt{2}\,N}~,
\label{CUloc}
\end{equation}
which, apart from the factor of $\sqrt{2}$ due to the $\mathbb{Z}_2$ orbifold, is the same expression of the 
$\mathcal{N}=4$ SYM theory \cite{Lee:1998bxa}. 

On the contrary, the correlators
with twisted operators depend on $\lambda$, as one can see already at 
the first perturbative order by using Feynman diagrams (see \cite{Billo:2019fbi,Galvagno:2020cgq,Billo:2021rdb}). Exploiting the matrix model formulation, one can go to very 
high orders in perturbation theory and, with limited computational effort, generate long series in $\lambda$. 
As shown in \cite{Billo:2021rdb} for the 2-point correlators, these series can be resummed in closed-form in terms of the matrix $\mathsf{X}$, thus obtaining the
$\lambda$-dependence beyond perturbation theory. From the asymptotic 
behavior of $\mathsf{X}$ for $\lambda\to\infty$ \cite{Beccaria:2021hvt,Beccaria:2021vuc}, we can then determine the 
coefficients $G_{T_k}$ at strong coupling. In \cite{Billo:2022xas} these methods have been generalized to the 3-point correlators in an orientifold model. Building on these
results, we have further extended these calculations to quiver theories \cite{Billo:2022fnb}
and managed to obtain an analytic expression of $G_{U_k T_\ell \overbar{T}_{p}}$
for any value of $\lambda$ by resumming the perturbative expansions in terms of the matrix $\mathsf{X}$ which can be extrapolated to strong coupling.  For the two-node quiver of Fig.\,\ref{fig:model},
our findings can be summarized as
\begin{equation}
G_{T_k}
=\begin{cases}
\mathcal{G}_k~~\mbox{for}~\lambda\to 0~,\\[2mm]
\frac{4\pi^2\,k\,(k-1)}{\lambda}\,\mathcal{G}_k~~\mbox{for}~\lambda\to \infty~,
\end{cases}
\label{GTloc}
\end{equation}
and
\begin{equation}
G_{U_k T_\ell \overbar{T}_{p}} 
=\begin{cases}
\mathcal{G}_{k,\ell,p}~~\mbox{for}~\lambda\to 0~,\\[2mm]
\frac{4\pi^2\,(\ell-1)\,(p-1)}{\lambda}\,\mathcal{G}_{k,\ell,p}~~\mbox{for}~\lambda\to \infty~,
\end{cases}
\label{GUTTloc}
\end{equation}
where $\mathcal{G}_k$ and $\mathcal{G}_{k,\ell,p}$ are defined in (\ref{GUloc}) and (\ref{GUUUloc}).
From these expressions, it follows that 
\begin{equation}
C_{U_k T_\ell \overbar{T}_{p}}
=\begin{cases}
\frac{\sqrt{k\,\ell\,p}\phantom{\big|}}{\sqrt{2}\,N}~~\mbox{for}~\lambda\to 0~,\\[2mm]
\frac{\sqrt{k\,(\ell-1)\,(p-1)}\phantom{\big|}}{\sqrt{2}\,N}~~\mbox{for}~\lambda\to \infty~.
\end{cases}
\label{CTloc}
\end{equation}
\begin{figure}[ht]
	\includegraphics[width=0.9\columnwidth]{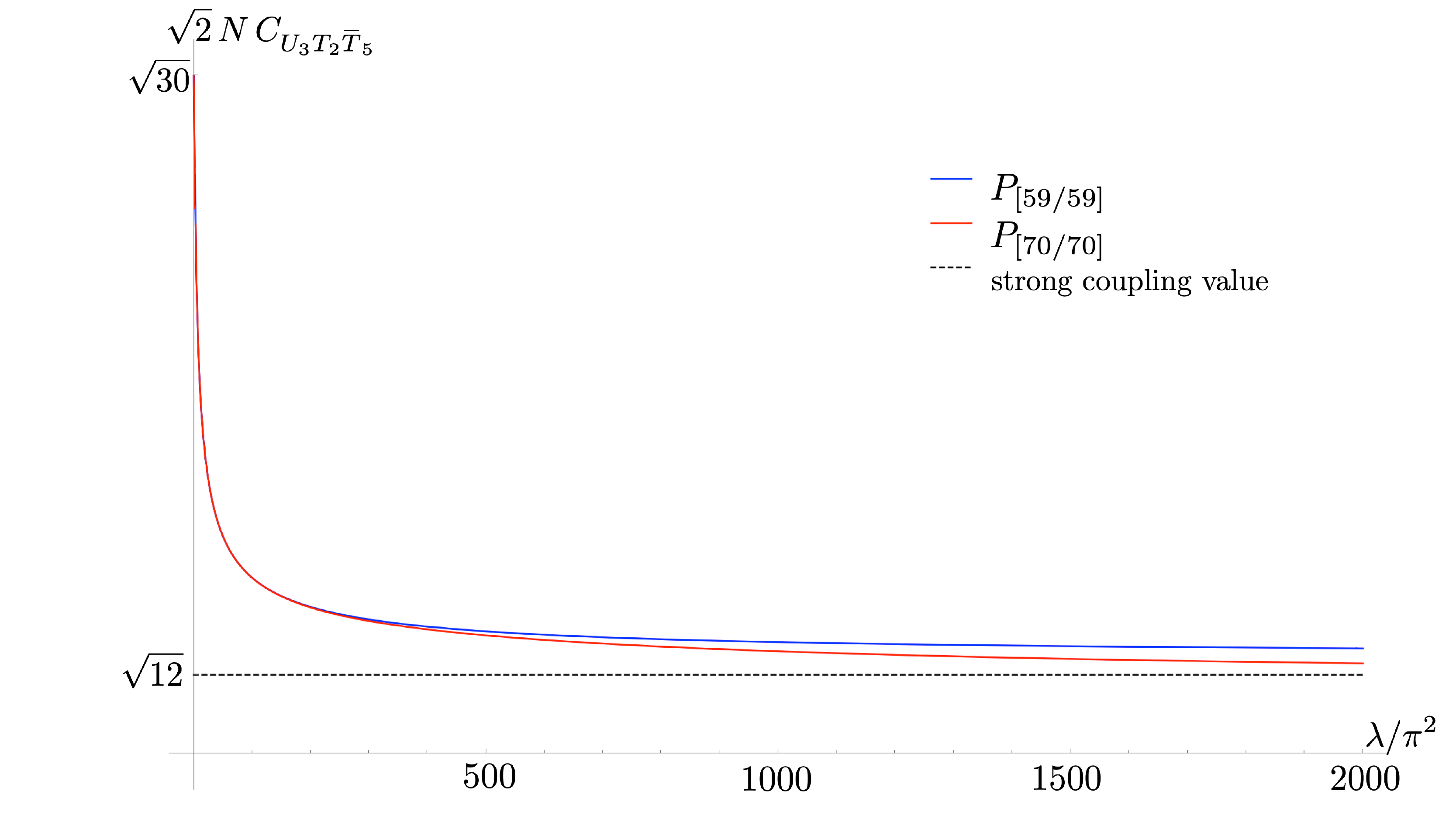}
	\caption{\label{fig:Cutt}Plot of the Pad\'e approximations of degree 59 (blue line) and
	70 (red line) of the structure constant $C_{U_3T_2\overbar{T}_{5}}$. For larger values of $\lambda$ they tend towards the predicted strong coupling value (dashed black line).}
\end{figure}
We emphasize that our methods allow to evaluate the structure constants for \emph{all} values of $\lambda$, and not only in the asymptotic regimes. As an example, in Fig.\,\ref{fig:Cutt}
we show how the structure constant $C_{U_3T_2\overbar{T}_{5}}$ varies in a region of intermediate values of $\lambda$, where we have done two Pad\'e resummations of the perturbative series.

\section{Holographic derivation}
\label{sec:holo}
We now derive the structure constants (\ref{CUUUdef}) and (\ref{CUTTdef}) at strong coupling using the AdS/CFT correspondence \cite{Maldacena:1997re,Gubser:1998bc,Witten:1998qj}. Since we are interested in the large-$N$ results, we can work in the classical supergravity approximation. However, it is useful to start from the string set-up.

We consider the  $\mathbb{Z}_2$ orbifold projection from a stack of $2N$ regular D3-branes in Type II B string theory that engineer a $\mathcal{N}=4$ SYM theory with gauge group SU($2N$). Breaking this configuration into two stacks of $N$ fractional D3-branes located at the orbifold fixed-point \cite{Douglas:1996sw}, we obtain the quiver theory of Fig.\,\ref{fig:model}.

The fractional D3-branes are soliton configurations emitting the metric and a 4-form potential $C_{4}$ with a self-dual field strength, together with the scalars
$b$ and $c$ corresponding to the wrapping of the 2-forms $B_{2}$ and $C_{2}$ around the exceptional 2-cycle of the orbifold 
\footnote{We recall that in presence of fractional D3-branes the dilaton and the axion are constant.}.
We therefore have an untwisted sector comprising the metric and the 4-form $C_{4}$, which propagate in ten dimensions, and a twisted sector with the scalars $b$ and $c$, which propagate in the six-dimensional space defined at the orbifold fixed point. In the near horizon limit these spaces become, respectively, 
AdS$_5\times S^5/\mathbb{Z}_2$ and AdS$_5\times S^1$ \cite{Kachru:1998ys,Gukov:1998kk}.

The dynamics of the untwisted fields is governed by the
equations of Type II B supergravity in AdS$_5\times S^5/\mathbb{Z}_2$
with the field-strength of $C_4$ proportional to the volume form
of the AdS$_5$ space.
The fluctuations about this background are
bulk fields
dual to the
untwisted operators of the quiver theory. 
We set \footnote{Our conventions are the following: ten-dimensional indices are denoted by Latin letters $m,n,\ldots$; five-dimensional indices in the AdS$_5$ space are denoted by Greek letters from the middle part of the alphabet $\mu,\nu,\ldots$, whereas five-dimensional indices along 
the 5-sphere are denoted by Greek letters from the beginning part of the alphabet $\alpha,\beta,\ldots.$.}
\begin{eqnarray}
&G_{mn}=g_{mn}+h_{mn} ~,\notag\\
&{C_4}_{\,m_1\ldots m_4}={c}_{m_1\ldots m_4}+{a}_{m_1\ldots m_4}~,
\label{back1}
\end{eqnarray}
where $g_{mn}$ and ${c}_{m_1\ldots m_4}$ are the background fields, while 
the fluctuations are as in \cite{Kim:1985ez,Lee:1998bxa}, namely
\begin{eqnarray}
&h_{\mu\nu}=h_{(\mu\nu)}^\prime-\frac{3}{25}\,h_2\,g_{\mu\nu}~~\mbox{with}~~
g^{\mu\nu}\,h_{(\mu\nu)}^\prime=0~,\notag\\
&h_{\alpha\beta}=h_{(\alpha\beta)}^\prime+\frac{1}{5}\,h_2\,g_{\alpha\beta}~~~\mbox{with}~~
g^{\alpha\beta}\,h_{(\alpha\beta)}^\prime=0~,\notag\\
&a_{\mu_1\mu_2\mu_3\mu_4}=-\,\epsilon_{\mu_1\mu_2\mu_3\mu_4\nu}\,\partial^\nu a~,\notag\\
&\!\!\!\! a_{\alpha_1\alpha_2\alpha_3\alpha_4}=\epsilon_{\alpha_1\alpha_2\alpha_3\alpha_4\beta}\,\partial^\beta a~.
\label{fluctuations}
\end{eqnarray}
We now expand the fluctuations in the spherical harmonics of $S^5$. Since we
are interested in those Kaluza-Klein (KK) modes which can couple to the fields of the twisted sector that are localized at the $\mathbb{Z}_2$ orbifold fixed point, we can restrict our attention to harmonics of the form
\begin{equation}
Y^k=\frac{1}{2^{\frac{|k|}{2}}}\,\mathrm{e}^{\mathrm{i}\,k\,\theta}\,\cos^{|k|}(\phi)
~~\mbox{for}~k\in\mathbb{Z}~,
\label{harmonics}
\end{equation}
where $\theta\in[0,2\pi]$ parametrizes the circle $S^1$
transverse to AdS$_5$ at the orbifold fixed point, and $\phi$ is one of the $S^5$ coordinates selected in such a way that $\phi=0$ corresponds to the 
fixed point.
The relevant expansions are
\begin{eqnarray}
h_2&=\sum_k h_{2,k}\,Y^k~,~~a=\sum_k a_{k}\,Y^k~,\notag\\[1mm]
&~~h_{(\mu\nu)}^\prime=\sum_k h_{(\mu\nu),k}^\prime\,Y^k~.
\label{expansions}
\end{eqnarray}
Doing the same analysis as in \cite{Kim:1985ez,Lee:1998bxa}, one can show that the combinations
\begin{equation}
s_k=\frac{1}{20(k+2)}\big[h_{2,k}-10\,(k+4)\,a_k\big]~,
\label{s}
\end{equation}
for $k>0$, and the complex conjugates $s_k^*$, 
correspond to KK excitations in AdS$_5$ with mass squared $m^2=k(k-4)$ that are
dual to the untwisted operators $U_k$ and $\overbar{U}_k$ of the quiver theory. Furthermore, as in \cite{Lee:1998bxa} we have
\begin{equation}
h_{(\mu\nu),k}^\prime=\frac{4\,\nabla_{(\mu}\nabla_{\nu)}s_k}{k+1}~,\quad
h_{(\mu\nu),-k}^\prime=\frac{4\,\nabla_{(\mu}\nabla_{\nu)}s_k^*}{k+1}
\label{hs}
\end{equation}
for $k>0$. The linearized equations for $s_k$ and
$s_k^*$ can be derived from the quadratic action
\begin{eqnarray}
S_{\mathrm{untw}}&=\displaystyle{\frac{4\,(2N)^2}{(2\pi)^5}}\!
\int_{\mathrm{AdS}_5}\!\!\!\!\!\!d^{5}z\,\sqrt{g}\,\,\sum_{k>0}A_k\,\Big[\,\partial s_k^*\cdot\partial
s_k\notag\\
&\qquad\qquad+k(k-4)\,s_k^*\,s_k\,\Big]\,\displaystyle{\frac{\pi^3}{2}}~,
\label{Suntw}
\end{eqnarray}
where the prefactor is the rewriting of the gravitational constant
using the parameters of the quiver theory in units where the radius of AdS$_5$ is set to one. The last factor of $\pi^3/2$ is the volume of $S^5/\mathbb{Z}_2$, and finally
\begin{equation}
A_k=\Big[\frac{32 \,k\,(k-1)\,(k+2)}{k+1}\Big] \Big[\frac{2}{2^{k}\,(k+1)(k+2)}\Big]
\end{equation}
where the first bracket was derived in \cite{Lee:1998bxa} and the second bracket 
comes from the normalization of the spherical harmonics (\ref{harmonics}).

Let us now turn to the twisted sector. As shown in \cite{Gukov:1998kk},
the dynamics of the scalars $b$ and $c$ is described by
\begin{eqnarray}
S_6&=\displaystyle{\frac{1}{2\kappa_{6}^2} }\bigg[
\int\! \!d^{6}x\,\sqrt{G_{6}}\, \Big(\frac{1}{2}\,
\partial b \cdot\partial b+\frac{1}{2}\,\partial c \cdot\partial c\big)\notag\\
&\qquad\quad+ 4
\displaystyle{\int \!C_{4}\wedge \,d b \,\wedge \,dc\,}
\bigg]~,
\label{S6}
\end{eqnarray}
where we have adopted the conventions of \cite{Billo:2021rdb}, with $G_{6}$ being the determinant of the metric and
$2\kappa_{6}^2$ the gravitational constant. Assuming 
$ds^2=ds^2_{\mathrm{AdS}_5}+d\theta^2$, 
and expanding $b$ and $c$ in harmonics of $S^1$, namely
\begin{equation}
b=\sum_k\,b_k\,\mathrm{e}^{\mathrm{i}\,k\,\theta}~,\quad
c=\sum_k\,c_k\,\mathrm{e}^{\mathrm{i}\,k\,\theta}~,
\end{equation}
one can show \cite{Gukov:1998kk,Billo:2021rdb,Billo:2022fnb} that the combination
\begin{equation}
\eta_k=c_k-\mathrm{i}\,b_k
\label{eta}
\end{equation}
for $k>0$, and its complex conjugate $\eta_k^*$, are KK excitations in AdS$_5$ with mass-squared $m^2=k(k-4)$ which are dual to the twisted operators $T_k$ and $\overbar{T}_k$ of the quiver theory. Their dynamics is governed by the action
\begin{eqnarray}
S_{\mathrm{tw}}&=\displaystyle{\frac{4\,(2N)^2}{(2\pi)^3\,2\,\lambda}}
\int_{\mathrm{AdS}_5}\!\!\!\!d^{5}z\,\sqrt{g}\,\,\sum_{k>0}
\frac{1}{2}\,\Big[\,\partial \eta_k^*\cdot\partial
\eta_k\notag\\
&\qquad\qquad+k(k-4)\,\eta_k^*\,\eta_k\,\Big]\,2\pi~,
\label{Stw}
\end{eqnarray}
where the prefactor comes from the gravitational constant $1/2\kappa_6^2$ of the orbifold using the 
AdS/CFT dictionary, and the last factor of $2\pi$ is just the length of $S^1$.

The actions (\ref{Suntw}) and (\ref{Stw}) can be used to obtain the 2-point functions (\ref{UU})
and (\ref{TT}) at large $N$ and at strong coupling with the AdS/CFT methods \cite{Witten:1998qj}.
Using Eq.\,(17) of \cite{Freedman:1998tz} and the correction factor in Eq.\,(95),
from (\ref{Suntw}) we deduce that
\begin{equation}
G_{U_k}=\frac{4\,(2N)^2}{(2\pi)^5}\,A_k\,\frac{1}{\pi^2}\,\frac{\Gamma(k+1)}{\Gamma(k-2)}\,\frac{2(k-2)}{k}\,\frac{\pi^3}{2}~.
\label{2ptUnt}
\end{equation}
In a similar fashion, from (\ref{Stw}) we find
\begin{equation}
G_{T_k}=\frac{4\,(2N)^2}{(2\pi)^3\,2\,\lambda}\,\frac{1}{2}\,\frac{1}{\pi^2}\,\frac{\Gamma(k+1)}{\Gamma(k-2)}\,\frac{2(k-2)}{k}\,2\pi~.
\label{2ptTw}
\end{equation}
In writing these formulas we have not taken into account the possible presence of an arbitrary proportionality constant in the coupling between the quiver operators and the supergravity modes 
on the boundary of AdS$_5$ since, as shown in \cite{Lee:1998bxa}, this constant drops out in the normalized structure constants.

To compute the 3-point correlators we have to work out the cubic interactions of the KK modes. For the untwisted ones, we can rely again on the analysis of \cite{Lee:1998bxa} which, translated in our
notations, leads to
\begin{eqnarray}
S_{\mathrm{untw}}^\prime&=\displaystyle{
\frac{4\,(2N)^2}{(2\pi)^5}}
\int_{\mathrm{AdS}_5}\!\!\!\!d^{5}z\,\sqrt{g}
\sum_{k,\ell,p>0}\!\Big[V_{k \ell p}\,s_k\,s_\ell\,s_p^*
\notag\\
&\qquad\qquad\qquad\times \delta_{k+\ell-p,0}+\mathrm{c.c.}
\Big]\,\displaystyle{\frac{\pi^3}{2}}~,
\label{Suntw3}
\end{eqnarray}
where the cubic coupling $V_{k\ell p}$ can be read from Eq.s (3.39) and (3.40) of \cite{Lee:1998bxa}. From this action, using the AdS/CFT formulas of \cite{Freedman:1998tz}, we obtain
\begin{eqnarray}
&\hspace{-1.2cm}G_{U_k U_\ell \overbar{U}_p}
=\displaystyle{\frac{N^2}{2^{\frac{k+\ell+p}{2}-7}\,\pi^6}}\,\frac{k\,(k-1)\,(k-2)}{k+1}
\label{GUUUholo}\\
&\qquad\qquad\qquad\times\displaystyle{\frac{\ell\,(\ell-1)\,(\ell-2)}{\ell+1}\,\frac{p\,(p-1)\,(p-2)}{p+1}}~,
\notag
\end{eqnarray}
where the $\delta$-function imposing charge conservation is understood.
Combining (\ref{GUUUholo}) and (\ref{2ptUnt}), it 
follows that
\begin{equation}
C_{U_k U_\ell \overbar{U}_{p}} 
=\frac{\sqrt{k\,\ell\,p}\phantom{\big|}}{\sqrt{2}\,N}~,
\label{CUholo}
\end{equation}
in agreement with the localization result (\ref{CUloc}).

We now consider the twisted sector. In this case we have to expand the
twisted action (\ref{S6}) to first order in the fluctuations to obtain the couplings involving one untwisted mode and two twisted ones. Using (\ref{fluctuations}) and the relations (\ref{s}) and (\ref{hs}), up to a boundary term
we obtain
\begin{eqnarray}
S_{\mathrm{tw}}^\prime&=
\displaystyle{\frac{4\,(2N)^2}{(2\pi)^3\,2\,\lambda}}
\int_{\mathrm{AdS}_5}\!\!\!\!\!\!d^{5}z\,\sqrt{g}\!
\sum_{k,\ell,p>0}\frac{1}{2}\,\Big[W_{k\ell p}\,s_k\,\eta_\ell\,\eta_p^*
\notag\\
&\qquad\qquad\qquad\times \delta_{k+\ell-p,0}+\mathrm{c.c.}
\Big]\,2\pi~,
\label{Stw3}
\end{eqnarray}
where \cite{Billo:2022fnb}
\begin{eqnarray}
&\hspace{-1cm}W_{k \ell p}=-(k+\ell-p)\,(k+p-\ell)\\
&\qquad\qquad\qquad\times\displaystyle{\frac{(k+\ell+p-2)\,(k+\ell+p-4)}{2^{\frac{k}{2}}\,(k+1)}}
~.\notag
\end{eqnarray}
We observe that if one uses the $\delta$-function that imposes charge conservation, this cubic coupling vanishes. This is a well-known feature of all couplings related to extremal correlators
\cite{DHoker:1999jke,Rastelli:2017udc}, and is not in contradiction 
with the fact that the final correlators are non-vanishing. Indeed, the zero in the coupling coefficient is compensated by a pole in the cubic Witten diagram of the 3-point function, so that the 
product yields a finite result. This can be clearly seen 
\cite{Lee:1998bxa,DHoker:1999jke,Rastelli:2017udc} if one imposes
the charge-conserving $\delta$-function only at the end, as we are going to do
\footnote{If one wishes to impose $p=k+\ell$, one has to carefully consider the boundary term, 
which otherwise does not contribute. However, as shown in \cite{DHoker:1999jke} these two approaches lead to the same results.}. With this understanding, using the AdS/CFT formulas of 
\cite{Freedman:1998tz}, we obtain
\begin{eqnarray}
&G_{U_k T_\ell \overbar{T}_p}
=\displaystyle{\frac{N^2}{2^{\frac{k}{2}-3}\,\pi^6\,\lambda}}\,\frac{k\,(k-1)\,(k-2)}{k+1}
\label{GUTTholo1}\\[1mm]
&\qquad\qquad\qquad\times (\ell-1)\,(\ell-2)\,(p-1)\,(p-2)~.
\notag
\end{eqnarray}
Then, from (\ref{2ptTw}) and (\ref{2ptUnt}), it follows that
\begin{equation}
C_{U_k T_\ell \overbar{T}_{p}}
=\frac{\sqrt{k\,(\ell-1)\,(p-1)}\phantom{\big|}}{\sqrt{2}\,N}~,
\label{CTholo}
\end{equation}
which confirms the localization result (\ref{CTloc}) at strong coupling.

\section{Conclusions}
By exploiting the power of supersymmetric localization we were able to obtain the exact $\lambda$-dependence of the structure constants of single-trace operators in the 2-node
quiver theory of Fig.\,\ref{fig:model} in the large-$N$ limit. 
When all operators are untwisted, the structure constants are $\lambda$-independent, like in 
the $\mathcal{N}=4$ SYM theory, but when two of the operators are twisted they depend on $\lambda$ in a highly non-trivial way. These twisted structure constants are therefore observables which, remarkably, can be followed from weak to strong coupling in an analytic way. The strong-coupling behavior of the structure constants predicted by localization is confirmed by a holographic calculation based on the AdS/CFT correspondence. This agreement can be seen either as a validation of the strong-coupling extrapolation of the localization results or, alternatively, as an explicit 
check of the AdS/CFT correspondence for a non-maximally supersymmetric theory
in four dimensions.
We finally mention that these results can be generalized to quiver theories with more than two nodes, as well as to orientifold models \cite{Billo:2022fnb}.

\begin{acknowledgments}
We thank F. Fucito, F. Galvagno and J.~F. Morales for useful discussions.
\end{acknowledgments}

\providecommand{\noopsort}[1]{}\providecommand{\singleletter}[1]{#1}%


\begin{thebibliography}{26}%
\makeatletter
\providecommand \@ifxundefined [1]{%
 \@ifx{#1\undefined}
}%
\providecommand \@ifnum [1]{%
 \ifnum #1\expandafter \@firstoftwo
 \else \expandafter \@secondoftwo
 \fi
}%
\providecommand \@ifx [1]{%
 \ifx #1\expandafter \@firstoftwo
 \else \expandafter \@secondoftwo
 \fi
}%
\providecommand \natexlab [1]{#1}%
\providecommand \enquote  [1]{``#1''}%
\providecommand \bibnamefont  [1]{#1}%
\providecommand \bibfnamefont [1]{#1}%
\providecommand \citenamefont [1]{#1}%
\providecommand \href@noop [0]{\@secondoftwo}%
\providecommand \href [0]{\begingroup \@sanitize@url \@href}%
\providecommand \@href[1]{\@@startlink{#1}\@@href}%
\providecommand \@@href[1]{\endgroup#1\@@endlink}%
\providecommand \@sanitize@url [0]{\catcode `\\12\catcode `\$12\catcode
  `\&12\catcode `\#12\catcode `\^12\catcode `\_12\catcode `\%12\relax}%
\providecommand \@@startlink[1]{}%
\providecommand \@@endlink[0]{}%
\providecommand \url  [0]{\begingroup\@sanitize@url \@url }%
\providecommand \@url [1]{\endgroup\@href {#1}{\urlprefix }}%
\providecommand \urlprefix  [0]{URL }%
\providecommand \Eprint [0]{\href }%
\providecommand \doibase [0]{https://doi.org/}%
\providecommand \selectlanguage [0]{\@gobble}%
\providecommand \bibinfo  [0]{\@secondoftwo}%
\providecommand \bibfield  [0]{\@secondoftwo}%
\providecommand \translation [1]{[#1]}%
\providecommand \BibitemOpen [0]{}%
\providecommand \bibitemStop [0]{}%
\providecommand \bibitemNoStop [0]{.\EOS\space}%
\providecommand \EOS [0]{\spacefactor3000\relax}%
\providecommand \BibitemShut  [1]{\csname bibitem#1\endcsname}%
\let\auto@bib@innerbib\@empty
\bibitem [{\citenamefont {Kachru}\ and\ \citenamefont
  {Silverstein}(1998)}]{Kachru:1998ys}%
  \BibitemOpen
  \bibfield  {author} {\bibinfo {author} {\bibfnamefont {S.}~\bibnamefont
  {Kachru}}\ and\ \bibinfo {author} {\bibfnamefont {E.}~\bibnamefont
  {Silverstein}},\ }\bibfield  {title} {\bibinfo {title} {{4-D conformal
  theories and strings on orbifolds}},\ }\href
  {https://doi.org/10.1103/PhysRevLett.80.4855} {\bibfield  {journal} {\bibinfo
   {journal} {Phys. Rev. Lett.}\ }\textbf {\bibinfo {volume} {80}},\ \bibinfo
  {pages} {4855} (\bibinfo {year} {1998})},\ \Eprint
  {https://arxiv.org/abs/hep-th/9802183} {arXiv:hep-th/9802183} \BibitemShut
  {NoStop}%
\bibitem [{\citenamefont {Gukov}(1998)}]{Gukov:1998kk}%
  \BibitemOpen
  \bibfield  {author} {\bibinfo {author} {\bibfnamefont {S.}~\bibnamefont
  {Gukov}},\ }\bibfield  {title} {\bibinfo {title} {{Comments on $\mathcal{N}=2$ AdS
  orbifolds}},\ }\href {https://doi.org/10.1016/S0370-2693(98)01005-3}
  {\bibfield  {journal} {\bibinfo  {journal} {Phys. Lett. B}\ }\textbf
  {\bibinfo {volume} {439}},\ \bibinfo {pages} {23} (\bibinfo {year} {1998})},\
  \Eprint {https://arxiv.org/abs/hep-th/9806180} {arXiv:hep-th/9806180}
  \BibitemShut {NoStop}%
  \bibitem [{\citenamefont {Zarembo}(2020)}]{Zarembo:2020tpf}%
  \BibitemOpen
  \bibfield  {author} {\bibinfo {author} {\bibfnamefont {K.}~\bibnamefont
  {Zarembo}},\ }\bibfield  {title} {\bibinfo {title} {{Quiver CFT at strong coupling}},\ }\href {https://doi.org/10.1007/JHEP06(2020)055}
   {\bibfield  {journal} {\bibinfo
  {journal} {J. High Energy Phys.}\ }\textbf {\bibinfo {volume} {06}},\
  \bibinfo {pages} {055}} (\bibinfo {year} {2020}),
\ \Eprint
  {https://arxiv.org/abs/2003.00993} {arXiv:2003.00993 [hep-th]} \BibitemShut
  {NoStop}%
\bibitem [{\citenamefont {Maldacena}(1998)}]{Maldacena:1997re}%
  \BibitemOpen
  \bibfield  {author} {\bibinfo {author} {\bibfnamefont {J.~M.}\ \bibnamefont
  {Maldacena}},\ }\bibfield  {title} {\bibinfo {title} {{The Large N limit of
  superconformal field theories and supergravity}},\ }\href
  {https://doi.org/10.1023/A:1026654312961} {\bibfield  {journal} {\bibinfo
  {journal} {Adv. Theor. Math. Phys.}\ }\textbf {\bibinfo {volume} {2}},\
  \bibinfo {pages} {231} (\bibinfo {year} {1998})},\ \Eprint
  {https://arxiv.org/abs/hep-th/9711200} {arXiv:hep-th/9711200} \BibitemShut
  {NoStop}%
\bibitem [{\citenamefont {Gubser}\ \emph {et~al.}(1998)\citenamefont {Gubser},
  \citenamefont {Klebanov},\ and\ \citenamefont {Polyakov}}]{Gubser:1998bc}%
  \BibitemOpen
  \bibfield  {author} {\bibinfo {author} {\bibfnamefont {S.~S.}\ \bibnamefont
  {Gubser}}, \bibinfo {author} {\bibfnamefont {I.~R.}\ \bibnamefont
  {Klebanov}},\ and\ \bibinfo {author} {\bibfnamefont {A.~M.}\ \bibnamefont
  {Polyakov}},\ }\bibfield  {title} {\bibinfo {title} {{Gauge theory
  correlators from noncritical string theory}},\ }\href
  {https://doi.org/10.1016/S0370-2693(98)00377-3} {\bibfield  {journal}
  {\bibinfo  {journal} {Phys. Lett. B}\ }\textbf {\bibinfo {volume} {428}},\
  \bibinfo {pages} {105} (\bibinfo {year} {1998})},\ \Eprint
  {https://arxiv.org/abs/hep-th/9802109} {arXiv:hep-th/9802109} \BibitemShut
  {NoStop}%
\bibitem [{\citenamefont {Witten}(1998)}]{Witten:1998qj}%
  \BibitemOpen
  \bibfield  {author} {\bibinfo {author} {\bibfnamefont {E.}~\bibnamefont
  {Witten}},\ }\bibfield  {title} {\bibinfo {title} {{Anti-de Sitter space and
  holography}},\ }\href {https://doi.org/10.4310/ATMP.1998.v2.n2.a2} {\bibfield
   {journal} {\bibinfo  {journal} {Adv. Theor. Math. Phys.}\ }\textbf {\bibinfo
  {volume} {2}},\ \bibinfo {pages} {253} (\bibinfo {year} {1998})},\ \Eprint
  {https://arxiv.org/abs/hep-th/9802150} {arXiv:hep-th/9802150} \BibitemShut
  {NoStop}%
\bibitem [{Note1()}]{Note1}%
  \BibitemOpen
  \bibinfo {note} {This Letter contains the main ideas and results, while the
  derivation and the extension to quiver theories with an arbitrary number of
  nodes are contained in \cite {Billo:2022fnb}.}\BibitemShut {Stop}%
\bibitem [{\citenamefont {Baggio}\ \emph {et~al.}(2014)\citenamefont {Baggio},
  \citenamefont {Niarchos},\ and\ \citenamefont
  {Papadodimas}}]{Baggio:2014sna}%
  \BibitemOpen
  \bibfield  {author} {\bibinfo {author} {\bibfnamefont {M.}~\bibnamefont
  {Baggio}}, \bibinfo {author} {\bibfnamefont {V.}~\bibnamefont {Niarchos}},\
  and\ \bibinfo {author} {\bibfnamefont {K.}~\bibnamefont {Papadodimas}},\
  }\bibfield  {title} {\bibinfo {title} {{Exact correlation functions in
  $SU(2)~\mathcal N=2$ superconformal QCD}},\ }\href
  {https://doi.org/10.1103/PhysRevLett.113.251601} {\bibfield  {journal}
  {\bibinfo  {journal} {Phys. Rev. Lett.}\ }\textbf {\bibinfo {volume} {113}},\
  \bibinfo {pages} {251601} (\bibinfo {year} {2014})},\ \Eprint
  {https://arxiv.org/abs/1409.4217} {arXiv:1409.4217 [hep-th]} \BibitemShut
  {NoStop}%
\bibitem [{\citenamefont {Gerchkovitz}\ \emph {et~al.}(2017)\citenamefont
  {Gerchkovitz}, \citenamefont {Gomis}, \citenamefont {Ishtiaque},
  \citenamefont {Karasik}, \citenamefont {Komargodski},\ and\ \citenamefont
  {Pufu}}]{Gerchkovitz:2016gxx}%
  \BibitemOpen
  \bibfield  {author} {\bibinfo {author} {\bibfnamefont {E.}~\bibnamefont
  {Gerchkovitz}}, \bibinfo {author} {\bibfnamefont {J.}~\bibnamefont {Gomis}},
  \bibinfo {author} {\bibfnamefont {N.}~\bibnamefont {Ishtiaque}}, \bibinfo
  {author} {\bibfnamefont {A.}~\bibnamefont {Karasik}}, \bibinfo {author}
  {\bibfnamefont {Z.}~\bibnamefont {Komargodski}},\ and\ \bibinfo {author}
  {\bibfnamefont {S.~S.}\ \bibnamefont {Pufu}},\ }\bibfield  {title} {\bibinfo
  {title} {{Correlation Functions of Coulomb Branch Operators}},\ }\href
  {https://doi.org/10.1007/JHEP01(2017)103} {\bibfield  {journal} {\bibinfo
  {journal} {J. High Energy Phys.}\ }\textbf {\bibinfo {volume} {01}},\
  \bibinfo {pages} {103}} (\bibinfo {year} {2017}),\ \Eprint {https://arxiv.org/abs/1602.05971}
  {arXiv:1602.05971 [hep-th]} \BibitemShut {NoStop}%
\bibitem [{\citenamefont {Billo}\ \emph {et~al.}(2018)\citenamefont {Billo},
  \citenamefont {Fucito}, \citenamefont {Lerda}, \citenamefont {Morales},
  \citenamefont {Stanev},\ and\ \citenamefont {Wen}}]{Billo:2017glv}%
  \BibitemOpen
  \bibfield  {author} {\bibinfo {author} {\bibfnamefont {M.}~\bibnamefont
  {Billo}}, \bibinfo {author} {\bibfnamefont {F.}~\bibnamefont {Fucito}},
  \bibinfo {author} {\bibfnamefont {A.}~\bibnamefont {Lerda}}, \bibinfo
  {author} {\bibfnamefont {J.F.}\ \bibnamefont {Morales}}, \bibinfo {author}
  {\bibfnamefont {{\relax Ya}.~S.}\ \bibnamefont {Stanev}},\ and\ \bibinfo
  {author} {\bibfnamefont {C.}~\bibnamefont {Wen}},\ }\bibfield  {title}
  {\bibinfo {title} {{Two-point Correlators in N=2 Gauge Theories}},\ }\href
  {https://doi.org/10.1016/j.nuclphysb.2017.11.003} {\bibfield  {journal}
  {\bibinfo  {journal} {Nucl. Phys.}\ }\textbf {\bibinfo {volume} {B926}},\
  \bibinfo {pages} {427} (\bibinfo {year} {2018})},\ \Eprint
  {https://arxiv.org/abs/1705.02909} {arXiv:1705.02909 [hep-th]} \BibitemShut
  {NoStop}%
\bibitem [{\citenamefont {Beccaria}\ \emph {et~al.}(2020)\citenamefont
  {Beccaria}, \citenamefont {Bill\`o}, \citenamefont {Galvagno}, \citenamefont
  {Hasan},\ and\ \citenamefont {Lerda}}]{Beccaria:2020hgy}%
  \BibitemOpen
  \bibfield  {author} {\bibinfo {author} {\bibfnamefont {M.}~\bibnamefont
  {Beccaria}}, \bibinfo {author} {\bibfnamefont {M.}~\bibnamefont {Bill\`o}},
  \bibinfo {author} {\bibfnamefont {F.}~\bibnamefont {Galvagno}}, \bibinfo
  {author} {\bibfnamefont {A.}~\bibnamefont {Hasan}},\ and\ \bibinfo {author}
  {\bibfnamefont {A.}~\bibnamefont {Lerda}},\ }\bibfield  {title} {\bibinfo
  {title} {{$ \mathcal{N} $ = 2 Conformal SYM theories at large $N
  $}},\ }\href {https://doi.org/10.1007/JHEP09(2020)116} {\bibfield  {journal}
  {\bibinfo  {journal} {J. High Energy Phys.}\ }\textbf {\bibinfo {volume}
  {09}},\ \bibinfo {pages} {116}} (\bibinfo {year} {2020}),\ \Eprint {https://arxiv.org/abs/2007.02840}
  {arXiv:2007.02840 [hep-th]} \BibitemShut {NoStop}%
\bibitem [{\citenamefont {Galvagno}\ and\ \citenamefont
  {Preti}(2021)}]{Galvagno:2020cgq}%
  \BibitemOpen
  \bibfield  {author} {\bibinfo {author} {\bibfnamefont {F.}~\bibnamefont
  {Galvagno}}\ and\ \bibinfo {author} {\bibfnamefont {M.}~\bibnamefont
  {Preti}},\ }\bibfield  {title} {\bibinfo {title} {{Chiral correlators in $
  \mathcal{N} $ = 2 superconformal quivers}},\ }\href
  {https://doi.org/10.1007/JHEP05(2021)201} {\bibfield  {journal} {\bibinfo
  {journal} {J. High Energy Phys.}\ }\textbf {\bibinfo {volume} {05}},\
  \bibinfo {pages} {201}} (\bibinfo {year} {2021}),\ \Eprint {https://arxiv.org/abs/2012.15792}
  {arXiv:2012.15792 [hep-th]} \BibitemShut {NoStop}%
\bibitem [{\citenamefont {Beccaria}\ \emph {et~al.}(2021)\citenamefont
  {Beccaria}, \citenamefont {Bill\`o}, \citenamefont {Frau}, \citenamefont
  {Lerda},\ and\ \citenamefont {Pini}}]{Beccaria:2021hvt}%
  \BibitemOpen
  \bibfield  {author} {\bibinfo {author} {\bibfnamefont {M.}~\bibnamefont
  {Beccaria}}, \bibinfo {author} {\bibfnamefont {M.}~\bibnamefont {Bill\`o}},
  \bibinfo {author} {\bibfnamefont {M.}~\bibnamefont {Frau}}, \bibinfo {author}
  {\bibfnamefont {A.}~\bibnamefont {Lerda}},\ and\ \bibinfo {author}
  {\bibfnamefont {A.}~\bibnamefont {Pini}},\ }\bibfield  {title} {\bibinfo
  {title} {{Exact results in a $ \mathcal{N} $ = 2 superconformal gauge theory
  at strong coupling}},\ }\href {https://doi.org/10.1007/JHEP07(2021)185}
  {\bibfield  {journal} {\bibinfo  {journal} {J. High Energy Phys.}\ }\textbf
  {\bibinfo {volume} {07}},\ \bibinfo {pages} {185}} (\bibinfo {year} {2021}),\ \Eprint
  {https://arxiv.org/abs/2105.15113} {arXiv:2105.15113 [hep-th]} \BibitemShut
  {NoStop}%
\bibitem [{\citenamefont {Billo}\ \emph {et~al.}(2021)\citenamefont {Billo},
  \citenamefont {Frau}, \citenamefont {Galvagno}, \citenamefont {Lerda},\ and\
  \citenamefont {Pini}}]{Billo:2021rdb}%
  \BibitemOpen
  \bibfield  {author} {\bibinfo {author} {\bibfnamefont {M.}~\bibnamefont
  {Billo}}, \bibinfo {author} {\bibfnamefont {M.}~\bibnamefont {Frau}},
  \bibinfo {author} {\bibfnamefont {F.}~\bibnamefont {Galvagno}}, \bibinfo
  {author} {\bibfnamefont {A.}~\bibnamefont {Lerda}},\ and\ \bibinfo {author}
  {\bibfnamefont {A.}~\bibnamefont {Pini}},\ }\bibfield  {title} {\bibinfo
  {title} {{Strong-coupling results for $ \mathcal{N} $ = 2 superconformal
  quivers and holography}},\ }\href@noop {} {\bibfield  {journal} {\bibinfo
  {journal} {J. High Energy Phys.}\ }\textbf {\bibinfo {volume} {10}},\
  \bibinfo {pages} {161}} (\bibinfo {year} {2021}),\ \Eprint {https://arxiv.org/abs/2109.00559}
  {arXiv:2109.00559 [hep-th]} \BibitemShut {NoStop}%
\bibitem [{\citenamefont {Billo}\ \emph
  {et~al.}(2022{\natexlab{a}})\citenamefont {Billo}, \citenamefont {Frau},
  \citenamefont {Lerda}, \citenamefont {Pini},\ and\ \citenamefont
  {Vallarino}}]{Billo:2022xas}%
  \BibitemOpen
  \bibfield  {author} {\bibinfo {author} {\bibfnamefont {M.}~\bibnamefont
  {Billo}}, \bibinfo {author} {\bibfnamefont {M.}~\bibnamefont {Frau}},
  \bibinfo {author} {\bibfnamefont {A.}~\bibnamefont {Lerda}}, \bibinfo
  {author} {\bibfnamefont {A.}~\bibnamefont {Pini}},\ and\ \bibinfo {author}
  {\bibfnamefont {P.}~\bibnamefont {Vallarino}},\ }\href@noop {} {\bibinfo
  {title} {{Three-point functions in a $\mathcal{N}=2$ superconformal gauge
  theory and their strong-coupling limit}}} (\bibinfo {year}
  {2022}{\natexlab{a}}),\ \Eprint {https://arxiv.org/abs/2202.06990}
  {arXiv:2202.06990 [hep-th]} \BibitemShut {NoStop}%
\bibitem [{\citenamefont {Pestun}(2012)}]{Pestun:2007rz}%
  \BibitemOpen
  \bibfield  {author} {\bibinfo {author} {\bibfnamefont {V.}~\bibnamefont
  {Pestun}},\ }\bibfield  {title} {\bibinfo {title} {{Localization of gauge
  theory on a four-sphere and supersymmetric Wilson loops}},\ }\href
  {https://doi.org/10.1007/s00220-012-1485-0} {\bibfield  {journal} {\bibinfo
  {journal} {Commun. Math. Phys.}\ }\textbf {\bibinfo {volume} {313}},\
  \bibinfo {pages} {71} (\bibinfo {year} {2012})},\ \Eprint
  {https://arxiv.org/abs/0712.2824} {arXiv:0712.2824 [hep-th]} \BibitemShut
  {NoStop}%
\bibitem [{\citenamefont {Lee}\ \emph {et~al.}(1998)\citenamefont {Lee},
  \citenamefont {Minwalla}, \citenamefont {Rangamani},\ and\ \citenamefont
  {Seiberg}}]{Lee:1998bxa}%
  \BibitemOpen
  \bibfield  {author} {\bibinfo {author} {\bibfnamefont {S.}~\bibnamefont
  {Lee}}, \bibinfo {author} {\bibfnamefont {S.}~\bibnamefont {Minwalla}},
  \bibinfo {author} {\bibfnamefont {M.}~\bibnamefont {Rangamani}},\ and\
  \bibinfo {author} {\bibfnamefont {N.}~\bibnamefont {Seiberg}},\ }\bibfield
  {title} {\bibinfo {title} {{Three point functions of chiral operators in $D =
  4$, $\mathcal{N}=4$ SYM at large $N$}},\ }\href {https://doi.org/10.4310/ATMP.1998.v2.n4.a1}
  {\bibfield  {journal} {\bibinfo  {journal} {Adv. Theor. Math. Phys.}\
  }\textbf {\bibinfo {volume} {2}},\ \bibinfo {pages} {697} (\bibinfo {year}
  {1998})},\ \Eprint {https://arxiv.org/abs/hep-th/9806074}
  {arXiv:hep-th/9806074 [hep-th]} \BibitemShut {NoStop}%
\bibitem [{\citenamefont {Billo}\ \emph {et~al.}(2019)\citenamefont {Billo},
  \citenamefont {Galvagno},\ and\ \citenamefont {Lerda}}]{Billo:2019fbi}%
  \BibitemOpen
  \bibfield  {author} {\bibinfo {author} {\bibfnamefont {M.}~\bibnamefont
  {Billo}}, \bibinfo {author} {\bibfnamefont {F.}~\bibnamefont {Galvagno}},\
  and\ \bibinfo {author} {\bibfnamefont {A.}~\bibnamefont {Lerda}},\ }\bibfield
   {title} {\bibinfo {title} {{BPS Wilson loops in generic conformal $
  \mathcal{N} $ = 2 SU(N) SYM theories}},\ }\href
  {https://doi.org/10.1007/JHEP08(2019)108} {\bibfield  {journal} {\bibinfo
  {journal} {J. High Energy Phys.}\ }\textbf {\bibinfo {volume} {08}},\
  \bibinfo {pages} {108}} (\bibinfo {year} {2019}),\ \Eprint {https://arxiv.org/abs/1906.07085}
  {arXiv:1906.07085 [hep-th]} \BibitemShut {NoStop}%
\bibitem [{\citenamefont {Beccaria}\ \emph
  {et~al.}(2021{\natexlab{b}})\citenamefont {Beccaria}, \citenamefont {Dunne},\
  and\ \citenamefont {Tseytlin}}]{Beccaria:2021vuc}%
  \BibitemOpen
  \bibfield  {author} {\bibinfo {author} {\bibfnamefont {M.}~\bibnamefont
  {Beccaria}}, \bibinfo {author} {\bibfnamefont {G.~V.}\ \bibnamefont
  {Dunne}},\ and\ \bibinfo {author} {\bibfnamefont {A.~A.}\ \bibnamefont
  {Tseytlin}},\ }\bibfield  {title} {\bibinfo {title} {{BPS Wilson loop in $
  \mathcal{N} $ = 2 superconformal SU(N)
  \textquotedblleft{}orientifold\textquotedblright{} gauge theory and
  weak-strong coupling interpolation}},\ }\href
  {https://doi.org/10.1007/JHEP07(2021)085} {\bibfield  {journal} {\bibinfo
  {journal} {JHEP}\ }\textbf {\bibinfo {volume} {07}},\ \bibinfo {pages}
  {085}} (\bibinfo {year} {2021}),\ \Eprint {https://arxiv.org/abs/2104.12625} {arXiv:2104.12625
  [hep-th]} \BibitemShut {NoStop}%
\bibitem [{\citenamefont {Billo}\ \emph
  {et~al.}(2022{\natexlab{b}})\citenamefont {Billo}, \citenamefont {Frau},
  \citenamefont {Lerda}, \citenamefont {Pini},\ and\ \citenamefont
  {Vallarino}}]{Billo:2022fnb}%
  \BibitemOpen
  \bibfield  {author} {\bibinfo {author} {\bibfnamefont {M.}~\bibnamefont
  {Billo}}, \bibinfo {author} {\bibfnamefont {M.}~\bibnamefont {Frau}},
  \bibinfo {author} {\bibfnamefont {A.}~\bibnamefont {Lerda}}, \bibinfo
  {author} {\bibfnamefont {A.}~\bibnamefont {Pini}},\ and\ \bibinfo {author}
  {\bibfnamefont {P.}~\bibnamefont {Vallarino}},\ }\bibfield  {title} {\bibinfo {title} {{Localization vs holography in $4d$ $
  \mathcal{N} =2$ quiver theories}},\ }\href
  {https://doi.org/10.48550/arXiv.2207.08846} (\bibinfo {year} {2022}),\ \Eprint {https://arxiv.org/abs/2207.08846} {arXiv:2207.08846
  [hep-th]} \BibitemShut {NoStop}%
\bibitem [{\citenamefont {Douglas}\ and\ \citenamefont
  {Moore}(1996)}]{Douglas:1996sw}%
  \BibitemOpen
  \bibfield  {author} {\bibinfo {author} {\bibfnamefont {M.~R.}\ \bibnamefont
  {Douglas}}\ and\ \bibinfo {author} {\bibfnamefont {G.~W.}\ \bibnamefont
  {Moore}},\ }\href@noop {} {\bibinfo {title} {{D-branes, quivers, and ALE
  instantons}}} (\bibinfo {year} {1996}),\ \Eprint
  {https://arxiv.org/abs/hep-th/9603167} {arXiv:hep-th/9603167} \BibitemShut
  {NoStop}%
\bibitem [{Note2()}]{Note2}%
  \BibitemOpen
  \bibinfo {note} {We recall that in presence of fractional D3-branes the
  dilaton and the axion are constant.}\BibitemShut {Stop}%
\bibitem [{Note3()}]{Note3}%
  \BibitemOpen
  \bibinfo {note} {Our conventions are the following: ten-dimensional indices
  are denoted by Latin letters $m,n,\ldots $; five-dimensional indices
  in the AdS$_5$ space are denoted by Greek letters from the middle part of the
  alphabet $\mu ,\nu ,\ldots $, whereas five-dimensional indices along
  the 5-sphere are denoted by Greek letters from the beginning part of the
  alphabet $\alpha ,\beta , \ldots .$}\BibitemShut {Stop}%
\bibitem [{\citenamefont {Kim}\ \emph {et~al.}(1985)\citenamefont {Kim},
  \citenamefont {Romans},\ and\ \citenamefont {van
  Nieuwenhuizen}}]{Kim:1985ez}%
  \BibitemOpen
  \bibfield  {author} {\bibinfo {author} {\bibfnamefont {H.~J.}\ \bibnamefont
  {Kim}}, \bibinfo {author} {\bibfnamefont {L.~J.}\ \bibnamefont {Romans}},\
  and\ \bibinfo {author} {\bibfnamefont {P.}~\bibnamefont {van
  Nieuwenhuizen}},\ }\bibfield  {title} {\bibinfo {title} {{The Mass Spectrum
  of Chiral $N=2$ $D=10$ Supergravity on $S^5$}},\ }\href
  {https://doi.org/10.1103/PhysRevD.32.389} {\bibfield  {journal} {\bibinfo
  {journal} {Phys. Rev. D}\ }\textbf {\bibinfo {volume} {32}},\ \bibinfo
  {pages} {389} (\bibinfo {year} {1985})}\BibitemShut {NoStop}%
\bibitem [{\citenamefont {Freedman}\ \emph {et~al.}(1999)\citenamefont
  {Freedman}, \citenamefont {Mathur}, \citenamefont {Matusis},\ and\
  \citenamefont {Rastelli}}]{Freedman:1998tz}%
  \BibitemOpen
  \bibfield  {author} {\bibinfo {author} {\bibfnamefont {D.~Z.}\ \bibnamefont
  {Freedman}}, \bibinfo {author} {\bibfnamefont {S.~D.}\ \bibnamefont
  {Mathur}}, \bibinfo {author} {\bibfnamefont {A.}~\bibnamefont {Matusis}},\
  and\ \bibinfo {author} {\bibfnamefont {L.}~\bibnamefont {Rastelli}},\
  }\bibfield  {title} {\bibinfo {title} {{Correlation functions in the
  CFT$_d$/AdS$_{d+1}$ correspondence}},\ }\href
  {https://doi.org/10.1016/S0550-3213(99)00053-X} {\bibfield  {journal}
  {\bibinfo  {journal} {Nucl. Phys. B}\ }\textbf {\bibinfo {volume} {546}},\
  \bibinfo {pages} {96} (\bibinfo {year} {1999})},\ \Eprint
  {https://arxiv.org/abs/hep-th/9804058} {arXiv:hep-th/9804058} \BibitemShut
  {NoStop}%
\bibitem [{\citenamefont {D'Hoker}\ \emph {et~al.}(1999)\citenamefont
  {D'Hoker}, \citenamefont {Freedman}, \citenamefont {Mathur}, \citenamefont
  {Matusis},\ and\ \citenamefont {Rastelli}}]{DHoker:1999jke}%
  \BibitemOpen
  \bibfield  {author} {\bibinfo {author} {\bibfnamefont {E.}~\bibnamefont
  {D'Hoker}}, \bibinfo {author} {\bibfnamefont {D.~Z.}\ \bibnamefont
  {Freedman}}, \bibinfo {author} {\bibfnamefont {S.~D.}\ \bibnamefont
  {Mathur}}, \bibinfo {author} {\bibfnamefont {A.}~\bibnamefont {Matusis}},\
  and\ \bibinfo {author} {\bibfnamefont {L.}~\bibnamefont {Rastelli}},\ }\href
  {https://doi.org/10.1142/9789812793850_0020} {\bibinfo {title} {{Extremal
  correlators in the AdS/CFT correspondence}}} (\bibinfo {year} {1999}),\
  \Eprint {https://arxiv.org/abs/hep-th/9908160} {arXiv:hep-th/9908160}
  \BibitemShut {NoStop}%
\bibitem [{\citenamefont {Rastelli}\ and\ \citenamefont
  {Zhou}(2018)}]{Rastelli:2017udc}%
  \BibitemOpen
  \bibfield  {author} {\bibinfo {author} {\bibfnamefont {L.}~\bibnamefont
  {Rastelli}}\ and\ \bibinfo {author} {\bibfnamefont {X.}~\bibnamefont
  {Zhou}},\ }\bibfield  {title} {\bibinfo {title} {{How to Succeed at
  Holographic Correlators Without Really Trying}},\ }\href
  {https://doi.org/10.1007/JHEP04(2018)014} {\bibfield  {journal} {\bibinfo
  {journal} {J. High Energy Phys.}\ }\textbf {\bibinfo {volume} {04}},\
  \bibinfo {pages} {014}} (\bibinfo {year} {2018}),\ \Eprint {https://arxiv.org/abs/1710.05923}
  {arXiv:1710.05923 [hep-th]} \BibitemShut {NoStop}%
\bibitem [{Note4()}]{Note4}%
  \BibitemOpen
  \bibinfo {note} {If one wishes to impose $p=k+\ell $, one has to
  consider the boundary term, which otherwise does not contribute. These two approaches lead to the same  results \cite {DHoker:1999jke}.}\BibitemShut {Stop}%
\end{thebibliography}
\end{document}